\documentclass[11pt,aps,nofootinbib,amsfonts,superscriptaddress,showpacs,showkeys]{revtex4}

\usepackage{amsfonts,amssymb,amscd,amsmath}
\usepackage{graphicx}

\newcommand{\slashit}[1]{#1 \kern-.45em\slash}

\newcommand{\slashP}{P \kern-.65em\slash }

\begin{document}
\title{Deconfinement, naturalness and the nuclear-quark equation of state}
\author{Amir H. Rezaeian}
\email{Rezaeian@tphys.uni-heidelberg.de}
\affiliation{Institute for Theoretical Physics, University of Heidelberg,
Philosophenweg 19, D-69120 Heidelberg, Germany}
\date{\today}
\begin{abstract}
Baryon-loops vacuum contribution in renormalized models like the
Linear sigma model and the Walecka model give rise to large unnatural
interaction coefficients, indicating that the quantum vacuum is not
adequately described by long-range degrees of freedom. We extend such
models into nonrenormalizable class by introducing an ultraviolet
cutoff into the model definition and treat the Dirac-sea
explicitly. In this way, one can avoid unnaturalness. We calculate the
equation of state for symmetric nuclear matter at zero temperature in
a modified $\sigma-\omega$ model. We show that the strong attraction
originating from the Dirac-sea softens the nuclear matter equation of
state and generates a vacuum with dynamically broken symmetry. In this model the
vector-meson is important for the description of normal nuclear
matter, but it obstructs the chiral phase transition.  We investigate
the chiral phase transition in this model by incorporating
deconfinement at high density.  A first-order quark deconfinement is
simulated by changing the active degrees of freedom from nucleons to
quarks at high density. We show that the chiral phase transition is
first-order when quark decouples from the vector-meson and coincides
with the deconfinement critical density.
\end{abstract}
\maketitle
\date{\today}


\section{Introduction}
The equation of state for strongly interacting matter at finite
density is needed for understanding of neutron stars \cite{h1} and
heavy ion collision phenomenology \cite{h2}. One of the most
interesting open questions for such system is the identification of
the appropriate degrees of freedom to describe the different phases of
matter. One expects that effective degrees of freedom change from
baryons at low density to quarks and gluons at high density.
Therefore, at high density nuclear matter undergoes a phase transition
to deconfined quarks and gluons and chiral symmetry is
restored. Our current knowledge about this transition is still very
rudimentary since QCD lattice computation cannot yet be performed at
large density. In this paper we attempt to describe the first order
liquid-gas phase transition of nuclear matter, the chiral phase transition, and the
transition to quark matter in a unified framework. For this purpose,
we work within an effective chiral linear sigma model coupled to
nucleons and quarks.

The chiral linear sigma model and its extension have been successful
in explaining the low energy physics of mesons and nucleons in vacuum
due to utilizing the concept of spontaneous chiral symmetry
breaking. However this model has been less
successful at finite density. Very early studies by Kerman and Miller \cite{h3} showed that the linear
sigma model (L$\sigma$N) model with nucleonic degrees of freedom in
fact does not lead to saturating nuclear matter in mean field
approximation.  This failure is due to the fact that in chiral models
with Mexican hat potential, the medium effect moves the vacuum
effective potential towards smaller values. This implies a smaller
curvature and consequently the mass of the scalar field decreases. The scalar
field in the linear sigma model plays the role of the chiral partner of the
pion and the mediator of medium-range nucleon-nucleon attraction.
Therefore, as its mass becomes smaller the attraction between
nucleons becomes stronger. This simple effect may destroy the
stability of nuclear matter\footnote{Note that assignment of a single meson
serving in both roles does not nesscesaily lead to a consistent
picture.  For example, in the successful non-chiral Walecka model \cite{h4,h4-11}, the
scalar meson only plays the role of the mediator of mid-range
attraction between nucleons. One should also note that the chiral
symmetry realization of the linear sigma model is neither unique nor 
essential. For example, in a non-linear realization \cite{h5}, the pions field appears
alone and in vector manifestation scenario of Harada and Yamawaki \cite{h6}, the rho
meson is taken as chiral partner of the pions.}.

Boguta \cite{h7} showed, however, that a saturating normal ground
state in the L$\sigma$N model can be reproduced by introducing a
vector-scalar coupling in a chirally invariant way and generating a
mass to the vector-meson via spontaneous symmetry breaking. In
this approach the saturation stems from the cancellation of large
repulsive vector mesons and attractive scalar mesons which is
generated through spontaneous symmetry breaking.  However,
unacceptably large compression modulus of nuclear matter (about
$K\simeq 650$ MeV) still remain difficult to overcome within this
approach. Another short-coming of this approach is the absence of
the chiral phase transition, since the effective nucleon mass tends to
grow at high density \cite{h7}. There has been some attempt to overcome
these difficulties by introducing an additional field, the dilaton
which is devised to effectively incorporate the trace anomaly in QCD
\cite{h8}. However, it was shown that a chirally restored phase in the presence
of vector-meson coupling to the dilaton is only possible at
unphysically high values of compressibility ($K\geq 1400$ MeV)
\cite{h8}. This paper is an attempt to resolve both problems, namely high
compressibility and the absence of abnormal phase. We will show that
inclusion of the Dirac-sea in a non-renormaliazble fashion reduces the
compressibility.  The chiral phase transition is expected to occur at
high density where the baryonic degrees of freedom are no longer
appropriate.  We show that the chiral restoration can be described by
incorporating the quark degrees of freedom into the L$\sigma$N model
at high density.  We simulate the quark deconfinement at finite
density by changing the active degrees of freedom from nucleons to
quarks at high density.  This leads to a first order deconfinement
phase transition, in accordance with indications from earlier studies
\cite{h9}.  In our simple model quark
deconfinement and chiral restoration are interconnected. A general
outcome of our analysis concerns the crucial importance of
deconfinement for understanding the chiral restoration at finite
density.

Initially, the quantum hydrodynamical model was based on
renormalizable field theory. Therefore, a systematic inclusion of
vacuum loops was necessary in order to accommodate the response of a
filled Dirac-sea to the presence of valence nucleons. By
renormalizability requirement one could then absorb the vacuum loops
and redefine the coefficients of the effective potential. However it
was shown that the resulting effective potential does not provide an
acceptable description of the properties of finite nuclei, and
inclusion of next-to-leading order loop expansion even worsens the
situation \cite{h10}. Furnstahl {\em et. al.} \cite{h11} showed also that one-baryon-loop vacuum
contribution in renormalized models like the L$\sigma$N model and the Walecka
model give rise to large {\em unnatural} coefficients based on ``naive
dimensional analysis'' proposed by Georgi and Manohar \cite{h12}, indicating that
the quantum vacuum is not described adequately by long-range degree of
freedoms.

One possible resolution to the {\em unnaturalness} has been
non-renormalizable effective field theory approach, where one includes
all possible terms consistent with the underlying symmetry, hoping
that short-range physics can be taken into account by non-linear
higher order interaction terms. In this way, there is no reference to
a Dirac sea of nucleons. In this paper, we take a radical approach and
introduce an ultraviolet cutoff $\Lambda$ into the definition of the
chiral sigma model. In our approach the cutoff is taken as new extra
parameter which needs to be fixed by some phenomenological
input. Therefore, our model by default becomes non-renormalizable,
similar to the Nambu-Jona-Lasinio (NJL) model \cite{h13}. In this way
by construction one does not need to absorb the vacuum loops into
interaction terms, and therefore the unnaturalness problem can be
avoided.  With the cutoff we can discard the short-range physics which
cannot be described by long-range degrees of freedom. In other words,
the sharp cutoff $\Lambda$ separates the active negative energy states
from the ones that does not contribute. In this way, we explicitly
retain the intuitive concept of Dirac sea as a counterpart of Fermi
sea. 

In the first part of the paper (section II) we study nuclear matter
properties within an extended L$\sigma$N model in present of explicit
Dirac-sea. In the second part of paper (section III), we incorporate
the quark degrees of freedom above deconfinement density. Having
calibrated the parameters of the model to the empirical nuclear matter
saturation, we investigate the implication of quark deconfinement in
the equation of state and in the description of the chiral
restoration. Some concluding remarks are given in section IV.

\section{Nuclear matter and naturalness}
We consider the chiral L$\sigma$N model with exact global $SU(2)\times
SU(2)$ symmetry, which contains a pseudoscalar coupling between pions
and nucleons, plus an auxiliary scalar field $\sigma$. We also include
a massive isoscalar vector field $\omega^{\mu}$,
\begin{equation}
\mathcal{L}_{n}=
\bar{\psi}_{n}[\gamma_{\mu}(i\partial^{\mu}-g_{vn}\omega^{\mu})-g_{s n}\left(\sigma+i\gamma_{5}\vec{\tau}\vec{\pi}\right)]\psi_{n}
+\mathcal{L}_{\sigma\pi}+\mathcal{L}_{\sigma\omega},  \label{n0}\
\end{equation}
where $\mathcal{L}_{\sigma\pi}$ and $\mathcal{L}_{\sigma\omega}$ are defined as
\begin{eqnarray}
\mathcal{L}_{\sigma\pi}&=&\frac{1}{2}\left(\partial_{\mu}\sigma\partial^{\mu}\sigma+\partial_{\mu}\vec{\pi}\partial^{\mu}\vec{\pi}\right)
+\frac{m_{s}^{2}}{2}(\sigma^{2}+\vec{\pi}^{2})+\frac{\lambda}{4}(\sigma^{2}+\vec{\pi}^{2})^{2}, \nonumber\\
\mathcal{L}_{\sigma\omega}&=&
-\frac{1}{4}F_{\mu\nu}F^{\mu\nu}+\frac{1}{2}g_{v}^{2}\omega^{\mu}\omega_{\mu}\left(\sigma^{2}+\vec{\pi}^{2}\right),
\label{l-s}\
\end{eqnarray}
and the field tensor is defined 
$F_{\mu\nu}=\partial_{\mu}\omega_{\nu}-\partial_{\nu}\omega_{\mu}$.
The nucleon mass $M_{N}$ and vector-meson mass $m_{v}$ at rest are
generated through spontaneous symmetry breaking by $\sigma$-field
\begin{equation}
M_{N}(\sigma)=g_{s n}\sigma, \hspace{2cm} m_{v}(\sigma)=g_{v}\sigma. \label{n1}
\end{equation}
For uniform nuclear matter, the ground state is obtained by filling
energy level with spin-isospin degeneracy $\gamma_{n}=4$ up to the
Fermi momentum $k_{F}$, where $k_{F}$ is related to the baryon density $\rho$ by
\begin{equation}
\rho=\frac{\gamma_{n}}{(2\pi)^{3}}\int^{k_{F}}_{0}d^{3}k.
\end{equation}
We employ the conventional mean-field approximation by expanding the
meson fields around their expectation values $\bar{\sigma}=\langle
\sigma\rangle$ and $\bar{\omega}^{0}=\langle \omega^{0}\rangle$ neglecting meson fluctuations. The mean values of the pion field and space component of the vector field vanish for a baryonic matter at rest due to symmetry. 
Having used the equation of motion for the $\bar{\omega}^{0}$ field, the energy density can be expressed as
\begin{eqnarray}
\Omega&=&\frac{g_{vn}^{2}\rho^{2}}{2m^{2}_{v}(\bar{\sigma})}+\frac{m_{s}^{2}}{2}\left(\bar{\sigma}^{2}-\bar{\sigma}^{2}_{0}\right)
+\frac{\lambda}{4}\left(\bar{\sigma}^{4}-\bar{\sigma}^{4}_{0}\right)
+\gamma_{n}\int^{k_{F}}_{0}\frac{d^{3}k}{(2\pi)^{3}}\sqrt{k^{2}+M^{2}_{N}(\bar{\sigma})}\nonumber\\
&-&\gamma_{n}\int^{\Lambda}_{0}\frac{d^{3}k}{(2\pi)^{3}}\left(\sqrt{k^{2}+M^{2}_{N}(\bar{\sigma})}-\sqrt{k^{2}+M^{2}_{N}(\bar{\sigma}_{0})}\right),\label{energy}\
\end{eqnarray}
where nucleon and vector masses are defined in Eq.~(\ref{n1}). The
$\bar{\sigma}_{0}$ denotes the mean-value of sigma field in vacuum.
The second line in Eq.~(\ref{energy}) describes the negative-energy
nucleon Dirac-sea contribution $\Omega_{vac}$ subtracted from its
corresponding value at zero density.  This term is divergent, we
regularize it by a sharp ultraviolet cutoff $\Lambda$. The linear
sigma model is renormalizable, therefore any cutoff dependence should
be removed by some kind of renormalization scheme in a such way that
the chiral symmetry of the model is preserved.  The vacuum expression
can be expanded in powers of $M_{N}/\Lambda$ which creates an infinite
series in terms of scalar mean-field (notice that the only first four
orders are divergent). By employing a renormalization scheme, one can
then partly absorb these terms into the coefficients of the effective
potential \cite{h15} (note also that this procedure is not unique),
\begin{equation}
\Omega_{vac}\simeq-\frac{\gamma_{n}}{(4\pi)^{2}}\left(M_{N}^{4}\ln\frac{M_{N}}{\mu}+\sum^{4}_{i=0}a_{i}(g_{s n}\sigma)^{i}\right), 
\label{pot}
\end{equation}
 where $\mu$ denotes the renormalization scale (which is normally
 chosen to be nucleon mass at zero density) and $a_{i}$ denotes the
 counterterms. The logarithmic term $M_{N}^{4}\ln\frac{M_{N}}{\mu}$,
 however cannot be removed in this fashion. The renormalizability
 requirement imposes very restrictive conditions on these highly
 nonlinear terms. It has been shown that such a renormalized theory is
 in apparent conflict with the {\em naturalness} property based on
 Georgi's naive dimensional analysis \cite{h12}. This indicates that
 the power counting will not be correct\footnote{Moreover, the
 resulting renormalized effective model fails to reproduce observed
 properties of finite nuclei, such as spin-orbit splitting, shell
 structures, charge densities, etc \cite{h0}.}.

The basic assumption of {\em naturalness} is that once the appropriate
dimensional scale has been extracted using ``naive dimensional
analysis'' proposed by Georgi and Manohar \cite{h12}, the remaining dimensionless
coefficients should all remain of order unity. If the naturalness assumption is valid, then the
effective Lagrangian can be truncated within a reasonable
error. Notice that there is no general proof of the naturalness property,
since we do not know how to derive effective hadronic Lagrangian from
QCD. Nevertheless, phenomenological studies support the
validity of naturalness and naive power counting rules.

The first attempt to make this model compatible with the naturalness
property has been to relax the renormalizability requirement and
include all non-linear terms which are allowed according to the
underlying symmetry \cite{h11}. Then, one can fix the parameters of
the model with some phenomenological input. In this fashion, no
reference to Dirac-sea is invoked and one retains only the valence
nucleon explicitly. All vacuum-loop effects will then be hidden in the
new interaction terms added to the original model. Here, we follow the
same line of idea, namely relaxing the renormalizability condition by
adding the ultraviolet cutoff $\Lambda$ to our model definition, but
we keep the concept of the Dirac-sea in a similar way to the
non-renormalizable NJL model \cite{h13}.  
Here, the Dirac-sea is explicitly kept without invoking any new ad hoc
interaction terms by expansion.

The in-medium scalar mean-field $\bar{\sigma}$
(and its corresponding value in vacuum $\bar{\sigma}_{0}$) or equivalently, the effective nucleon mass
$M_{N}(\bar{\sigma})$ ( and $M_{N}(\bar{\sigma}_{0})$ at zero density) is determined via the self-consistency condition by minimizing the energy-density $\Omega(\bar{\sigma})$
with respect to $\bar{\sigma}$,
\begin{equation}
-\frac{4g_{vn}^{2}k_{F}^{6}}{9\pi^{4}g^{2}_{v}\bar{\sigma}^{3}}+m_{s}^{2}\bar{\sigma}+\lambda\bar{\sigma}^{3}+
\frac{g_{s n}^{2}\bar{\sigma}}{\pi^{2}}\left[k_{F}E(k_{F})-\Lambda E(\Lambda)+g^{2}_{s n}\bar{\sigma}^{2}\ln\frac{\Lambda+E(\Lambda)}{k_{F}+E(k_{F})}\right]=0,\label{gap}
\end{equation}
where $E(k_{F})$ and $E(\Lambda)$ in the above expression are defined
by $E(k)=\sqrt{k^{2}+g_{s n}\bar{\sigma}^{2}}$. This non-linear
equation should be solved for every point of density associated with Fermi-momentum $k_{F}$. The
sigma mass is given by
\begin{equation}
m_{\sigma}^{2}=\frac{\partial^{2}\Omega}{\partial\bar{\sigma}^{2}}\label{sig}
\end{equation}
where the derivatives are evaluated at the scalar mean-field
$\bar{\sigma}$ solution of the gap equation (\ref{gap}).  In the presence
of the nucleon Dirac-sea, one does not need to constraint the form of
potential from outset with $m_{s}^{2}<0$ in order to ensure the
spontaneous symmetry breaking. The negative energy
Dirac-sea contribution can indeed generate a dynamical broken vacuum 
analogous to the NJL model. Due to spontaneous symmetry breaking
the scalar mean field acquires a nonzero vacuum expectation value which
is equal to the pion decay constant $f_{\pi}$. Here, the pions are the massless
Nambu-Goldstone bosons.

Our model contains six parameters $g_{s n}, g_{v}, m_{s}^{2},
\lambda, \Lambda$ and $g_{vn}$. In contrast to NJL type models, here
everything can be determined from the gap equation\footnote{Notice
that in general the gap equation may have many solutions, one has to
find out which solution minimizes the effective potential.}
(\ref{gap}). In the NJL model the meson masses and pion decay
should be solved as well since they are described as collective
quark-antiquark excitations, while in L$\sigma$N model they are
represented as dynamical fields. We take the empirical values for pion
decay $f_{\pi}=93$ MeV, nucleon mass in vacuum
$M_{N}(\bar{\sigma}_{0})=939$ MeV and vector-meson mass in vacuum
$m_{v}(\bar{\sigma}_{0})=783$ MeV. These choices fix $g_{s n}=M_{N}/f_{\pi}=10.1$ and $g_{v}=m_{v}/f_{\pi}=8.42$
via\footnote{The value of $g_{s n}$ differs from the experimental
value $g_{s n}=13.5$. This is a typical situation in chiral models with
$g_{A}=1$ where nucleons are taken stuctureless.} Eq.~(\ref{n1}).  The
vector field coupling $g_{vn}$ is treated as a free parameter and will
be adjusted in medium in such a way that the equation of state
reproduces nuclear-matter saturation properties, which we define as a
binding energy per nucleon $E/A=-15.75$ MeV at a density corresponding
to a Fermi momentum of $k_{F}=1.3~
\text{fm}^{-1}$.  The coupling $m_{s}^{2},\lambda$ and $\Lambda$ can be determined by requiring the value of the pion decay 
$f_{\pi}=93$ MeV and value of sigma meson mass defined in
Eq.~(\ref{sig}). In this way, one still cannot uniquely fix the value
of the cutoff. Therefore, for sake of generality we consider various
parameter sets $A1,A2$ and $B1-B3$ given in table I which are
determined with the above-mentioned procedure and all pass through the
empirical nuclear matter saturation point $(E_{B}/A=-15.75~\text{MeV},
\rho_{0}=0.148~ \text{fm}^{-3})$.  The empirical compression modulus
(or pressure) is not used in the fitting procedure. Generally,
increasing the cutoff shift the saturation point to higher density,
however for not very high cutoff (given in table I) one can adjust
other parameters in order to reproduce the empirical saturation
point. Parameter sets $A1$ and $A2$ are chosen with condition that
$m_{s}^{2}<0$, (i.e. with initial Mexican hat potential even without
inclusion of the Dirac-sea) while for parameter sets $B1$-$B3$ we have
$m_{s}^{2}>0$ (in this case, the Dirac-sea generates a Mexican hat
potential). For various parameter sets a sigma-meson of mass about
$810$ MeV is needed in order to reproduce the empirical saturation
point (see table 1).  The value of nucleon masses $M^{\star}_{N}$ at
the saturation point for various parameter sets are also given in
table I.

\begin{table}
\caption{The parameters $\Lambda$, $m_{s}^{2}$, $\lambda$ and $g_{vn}$ for sets $A1, A2$ and $B1-B3$ 
are given (see the text for details).  All parameter sets reproduce the empirical saturation point  $(E_{B}/A=-15.75~\text{MeV},
\rho_{0}=0.148~ \text{fm}^{-3})$.  The corresponding sigma mass
$m_{\sigma}$ in the vacuum and the resulting nucleon mass
$M_{N}(\bar{\sigma})|_{\rho_{0}}=M_{N}^{\star}$ and the
compressibility $K$ at saturation density are also
given.\label{tab:modpar}}
\begin{ruledtabular}
\begin{tabular}{llllll}
Parameter &set $A1$ & set $A2$ &set $B1$ & set $B2$ & set $B3$\\
\hline
$\Lambda$ (MeV) & 256 &300   &324   &444.2 &483.2\\
$m_{s}^{2} (\text{GeV}^{2})$ &-0.147 &-0.041    &0.026    &0.544 &0.77 \\
$\lambda$ & 31.5 &27.4   &25  &7.0 & 0\\
$g_{vn}$  & 6.69& 6.62   &6.54  &6.23 & 6.07\\
\hline
\hline
$m_{\sigma}(\bar{\sigma}_{0})$ (MeV) &806.5 & 805 &810.5   &810 & 817\\
$M_{N}^{\star}$(MeV)&757.2& 761.2 & 766.3  &781.3 &784.5\\
$K$ (MeV)&490&478&455&396&370\\
\end{tabular}
\end{ruledtabular}
\end{table}
On the right panel of Fig.~1, we show the mean-value of scalar field
$\bar{\sigma}$ with respect to density $\rho/\rho_{0}$.  As it is
observed as the density is increased the mean-value of scalar field
decreases. However, at high density above $2\rho_{0}$ the
curves (for different parameter sets) bend upward, indicating that the
chiral restoration can not be described in this model
\cite{h7,h8}. This is due to the fact that the model is
inadequate in short-distance physics while the meson degrees of
freedom $\sigma$ and $\omega$ account for nuclear interaction at short
distance.  Moreover, the vector-meson mass appears in the denominator
of the energy density Eq.~(\ref{energy}) and cannot vanish. On the
other hand, the vector-meson mass is locked to the scalar field by
chiral symmetry. Therefore, $\bar{\sigma}$ as well cannot vanish, our
numerical results reflect this fact. In the next section, we will
introduce a scenario in which the chiral restoration problem can be
resolved.

In Fig.~1 left panel, we show the density dependence of the energy per
baryon $E_{B}/A=\Omega/\rho-M_{N}(\bar{\sigma}_{0})$ calculated for
different parameter sets given in table I. The value of $\Omega$ is
calculated from Eq.~(\ref{energy}) by making use of the
self-consistency equation (\ref{gap}). It is observed from Fig.~1 that
as we increase the cutoff (increasing the contribution of the Dirac-sea)
the equation of state becomes softer. In order to find out quantitatively the
stiffness of the equation of state, we compute the compression
modulus. It is defined as
\begin{figure}[!tp]
       \centerline{\includegraphics[width=16 cm] {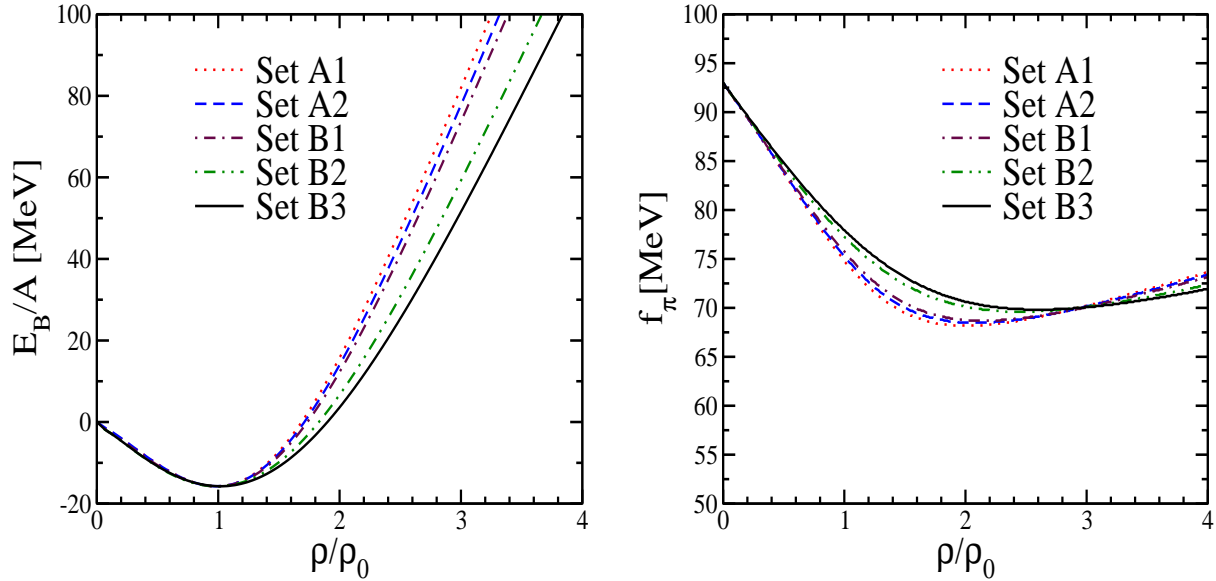}}
       \caption{On the right panel, we show in-medium pion decay as
       function of density $\rho/\rho_{0}$ (with nuclear matter
       density $\rho_{0}=0.15 fm^{-3}$) for various parameter sets
       given in table I. On the left panel, we show the binding
       energy per baryon $E_{B}/A$ as a function of density
       $\rho/\rho_{0}$.}
\end{figure}

\begin{equation}
K=9\frac{d^{2}\Omega}{d\rho^{2}}.
\end{equation}
The compression modulus at the saturation density for parameter set
$B3$ and $A1$ are $370$ MeV and $490$ MeV, respectively. The
compression modulus of other parameter sets lie between these two
values (see table I). These values are still bigger than the empirical
one $K=200-300$ MeV \cite{h16}. Nevertheless, it improves compared to
no-sea approximation where we have $K\simeq 650$ MeV \cite{h7}.  Note
that it has been shown by Prakash and Ainsworth \cite{h17} that the
empirical compressibility can be obtained when the vacuum-loops in a
renormalizible fashion are included. However, this results to
unacceptably large scalar sigma mass of about $1$ GeV and
unnaturalness of the effective Lagrangian as we already discussed. It
has been pointed out by Furnstahl and Serot \cite{h0} that the scalar
meson mass must be light enough in order to avoid strong fluctuations
in the charge density, which signal impending instabilities.


By increasing the cutoff, the value of the nucleon mass at saturation
density slightly increases. For the various parameter sets given in
table I, at saturation we have
$M_{N}^{\star}/M_{N}(\bar{\sigma}_{0})=0.81-0.84$ which is in good
agreement with analysis by Mahaux and Sartor \cite{h18}. 
\begin{figure}[!tp]
       \centerline{\includegraphics[width=16 cm] {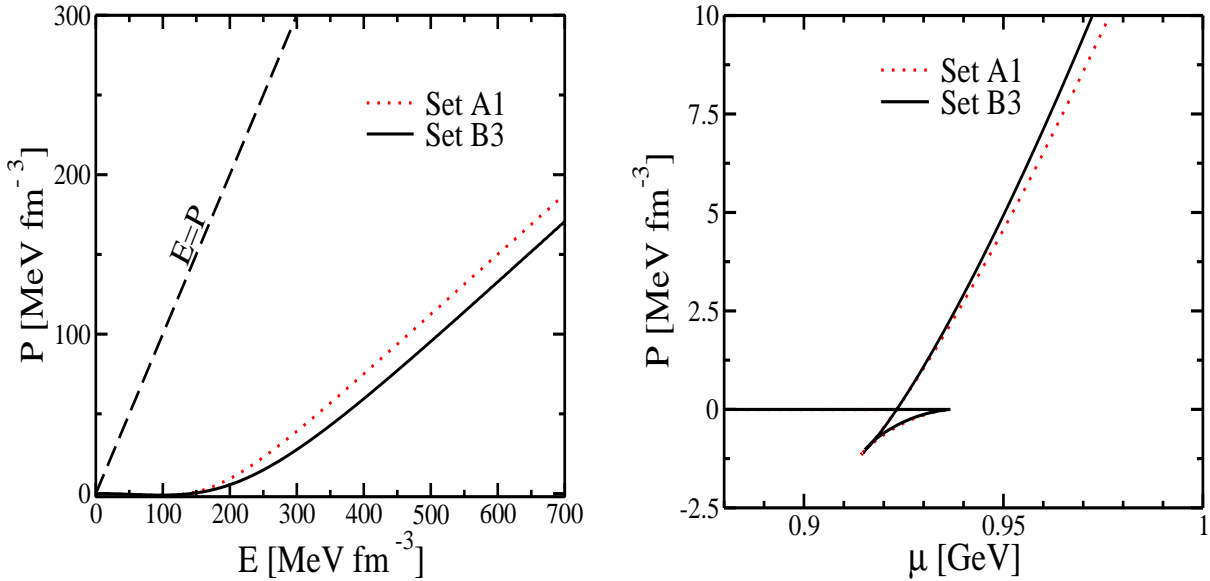}}
       \caption{ On the left, the pressure with respect to the energy is shown
       for parameter sets $A1$ and $B3$ given in table I. On the right side, we show the pressure as a function of
       chemical potential $\mu$ for the same parameter sets.}
\end{figure}

Next we compute the pressure $P$ at zero temperature as function of density
$\rho$, the baryonic chemical potential $\mu$ and energy. We use
thermodynamical identities
\begin{equation}
\mu=\frac{d\Omega}{d\rho}, \hspace{2cm} P=\mu\rho-\Omega=\rho^{2}\frac{d}{d\rho}\left(\frac{\Omega}{\rho}\right). \label{ther}
\end{equation}
In Fig.~2 left panel, we show the pressure $P$ as function of energy
$E=\Omega$ for set $A1$ and $B3$ (other parameter sets behave
similarly). The causal limit $E=P$ is also shown in the figure. It is
obvious that all the parameter sets studied here respect the causal
condition $\partial P/\partial E\leq 0$. Therefore,
non-renormalizibility in our model does not lead to conflicts with
causality. The pressure at vacuum and saturation density is zero
$P(\rho_{0})=0$ which gives a nontrivial constraint on the model
parameters. The phase transition becomes more apparent by computing
pressure as a function of chemical potential $\mu$. A first order
phase transition is manifested by appearance of several branches of
$P(\mu)$. In Fig.~2 right panel, the pressure with respect to chemical
potential for parameter sets $A1$ and $B3$ are shown. These curves
show the liquid-gas phase transition in nuclear matter. The first
order phase transition occurs at $\mu_{c}\simeq
M_{N}(\bar{\sigma}_{0})-15.75\simeq 923.25$ MeV for all various
parameter sets given in table I. At this point, the slope of curves
$P(\mu)$ exhibits a jump corresponding to difference of densities
between coexisting phases. At a fixed chemical potential only the
highest pressure corresponds to a stable phase.

Some remarks here are in order. One can observe that as we increase
the cutoff, i.e increasing the negative energy Dirac-sea contribution,
a stronger coupling $m_{s}^{2}$ (with positive sign) is needed, while
the coupling $\lambda$ subtantialy decreases (see table I). This can
be understood since the Dirac-sea contains many-body forces and in principle there
is no need to add arbitrary extra many-body forces in terms of
potential into system when the Dirac-sea is enhanced. On the other
hand, the nuclear matter saturation properties can be better described
(compressibility is reduced) with increasing cutoff in a acceptable
range. The Dirac-sea also generates a vacuum with dynamically broken
symmetry. Note that the exact connection between the chiral dynamics
and nuclear matter saturation mechanism is not yet completely
understood. Recently we have shown in a relativistic Faddeev approach
\cite{me} that nuclear matter saturation might be related to partial
chiral restoration \cite{h19}.

In the bosonized version of the NJL model, $\sigma^{4}$
self-interaction term is absent $\lambda=0$ (even in vacuum),
therefore our cutoff L$\sigma$N model calibrated to nuclear matter
saturation properties becomes very similar to the NJL model since we
have $\lambda\to 0$ as we increase the cutoff (see table I). It is
interesting to note as well that in mean field approximation, the
nuclear matter in the NJL model does not saturate without inclusion of
a vector-scalar fermion self-interaction \cite{h20}. Therefore the
same saturation mechanism is behind both models. Finally, as can be
noticed from table I, the acceptable range of cutoff is too low
$\Lambda\simeq 250-485$ MeV. A similar range for the cutoff has been
also reported in the NJL model with nucleonic degrees of freedom
\cite{h20,h21}.  This is in contrast with the case that the model is
defined in terms of quarks where the allowed range of the cutoff is
typically larger. This may indicate that by changing the relevant
degrees of freedom from nucleons to quarks, the model can be extended
to higher energy.

\section{Chiral restoration and deconfinement}
In this section we consider the implication of quark deconfinement
in the nuclear matter equation of state. We incorporate the
deconfinement effect by adding the quark dynamics to the nucleonic
Lagrangian $\mathcal{L}_{n}$ defined in Eq.~(\ref{n0}) and assume that
at a critical Fermi momentum $k_{C}$ (corresponding to a critical
density $\rho_{C}$), the baryonic degrees of freedom are replaced by
quark ones, while, keeping the description of scalar and pion fields
intact at higher density. We cannot derive the exact value of
deconfinement density $\rho_{C}$ in the context of mean field theory
where the internal structure of hadron is neglected, therefore we do
not constrain ourself to only one value for $\rho_{C}$ and consider
the implication of different values for $\rho_{C}$. The energy density
can be written
\begin{equation}
\Omega=\Omega_{n}+\Omega_{q}\theta(\rho-\rho_{C}), 
\end{equation}
where $\Omega_{n}$ is obtained from
$\mathcal{L}_{n}$ defined in Eq.~(\ref{n0}) and
$\Omega_{q}$ is obtained from the following quark Lagrangian  
\begin{equation}
\mathcal{L}_{q}=\bar{\psi}_{q}[\gamma_{\mu}i\partial^{\mu}-g_{s q}\left(\sigma+i\gamma_{5}\vec{\tau}\vec{\pi}\right)]\psi_{q}+\mathcal{L}_{\sigma\pi}+\mathcal{L}_{\sigma\omega},\label{l-q}
\end{equation}
where $\mathcal{L}_{\sigma\pi}$ and $\mathcal{L}_{\sigma\omega}$ are
defined in Eq.~(\ref{l-s}).  We do not couple the vector-meson field
directly to the quark field since the vector-meson in this model works
against chiral restoration phase transition at high
density. Therefore, we assume that the quark field is only coupled to
the scalar $\sigma$ and the pion $\vec{\pi}$ fields in chirally invariant
way. But we allow the vector-meson field to couple to the scalar field
through interaction term $\mathcal{L}_{\sigma\omega}$. Since quarks
are coupled to scalar field, the vector field is not completely
decoupled from quarks. The quark mass like the nucleon and vector meson
masses Eq.~(\ref{n1}) are generated through spontaneous symmetry
breaking via scalar field,
\begin{equation}
m_{q}(\sigma)=g_{s q}\sigma \label{qmass}.
\end{equation}
 The energy density $\Omega$ at finite density in
 mean field approximation as a function of Fermi momentum can be
 obtained as
\begin{eqnarray}
\Omega&=&\frac{2g_{vn}^{2}}{9\pi^{4}m^{2}_{v}(\bar{\sigma})}\left(k_{F}^{6} \theta(k_{C}-k_{F})+\mathcal{R} k_{C}^{6} \theta(k_{F}-k_{C})\right)
+\frac{m_{s}^{2}}{2}\left(\bar{\sigma}^{2}-\bar{\sigma}^{2}_{0}\right)
+\frac{\lambda}{4}\left(\bar{\sigma}^{4}-\bar{\sigma}^{4}_{0}\right)\nonumber\\
&+&\gamma_{n}\int^{k_{F}}_{0}\frac{d^{3}k}{(2\pi)^{3}}\sqrt{k^{2}+M^{2}_{N}(\bar{\sigma})}\theta(k_{C}-k_{F})
+\gamma_{n}\int^{k_{C}}_{0}\frac{d^{3}k}{(2\pi)^{3}}\sqrt{k^{2}+M^{2}_{N}(\bar{\sigma})}\theta(k_{F}-k_{C})\nonumber\\
&+&\gamma_{q}\int^{k_{F}}_{k_{C}}\frac{d^{3}k}{(2\pi)^{3}}\sqrt{k^{2}+m^{2}_{q}(\bar{\sigma})}\theta(k_{F}-k_{C})
-\gamma_{n}\int^{k_{C}}_{0}\frac{d^{3}k}{(2\pi)^{3}}\left(\sqrt{k^{2}+M^{2}_{N}(\bar{\sigma})}-\sqrt{k^{2}+M^{2}_{N}(\bar{\sigma}_{0})}\right)\nonumber\\
&-&\gamma_{q}\int^{\Lambda}_{k_{C}}\frac{d^{3}k}{(2\pi)^{3}}\left(\sqrt{k^{2}+m^{2}_{q}(\bar{\sigma})}-\sqrt{k^{2}+m^{2}_{q}(\bar{\sigma}_{0})}\right), \label{eqr}
\end{eqnarray}
where the quark mass $m_{q}$, the nucleon mass $M_{N}$ and the
vector-meson mass $m_{v}$ are defined in
Eqs.~(\ref{n1},\ref{qmass}). $\gamma_{q}=12$ and $\gamma_{n}=4$ are the
degeneracy factor for quarks and nucleon, respectively. We have also
included the corresponding Dirac-sea of the nucleons and the quarks
as in the previous section.  A priori we do not know the importance
of the vector-scalar interaction above deconfinement density. We
consider two cases above the critical deconfinement density
$\rho_{C}$: $g_{v}=0$ corresponding to totally decoupling of the
vector field from matter and $g_{v}\neq 0$ when the vector-scalar
interaction is present\footnote{In this case, the value of the
coupling $g_{v}$ is fixed by the empirical vector-meson mass in
vacuum.}. In the energy density equation (\ref{eqr}), the parameter
$\mathcal{R}$ can have two values $\mathcal{R}=0,1$ corresponding to the
solutions when $g_{v}=0$ and $g_{v}\neq 0$ above the critical
deconfinement density $\rho_{c}$, respectively. Note that the
parameter $\mathcal{R}$ was merely introduced in order to write both
solutions in one equation.

The Fermi momentum $k_{F}$ can be related to baryonic density by
\begin{equation}
\rho=\frac{2k^{3}_{F}}{3 \pi^{2}}. 
\end{equation}
The mean-value of scalar field
$\bar{\sigma}$ is determined by in-medium gap equation:
\begin{eqnarray}
&-&\frac{4g_{vn}^{2}}{9\pi^{4}g^{2}_{v}\bar{\sigma}^{3}}\left(k_{F}^{6}\theta(k_{C}-k_{F})+\mathcal{R}k_{C}^{6}\theta(k_{F}-k_{C})\right)
+m_{s}^{2}\bar{\sigma}+\lambda\bar{\sigma}^{3}+f_{q}(k_{C})-f_{q}(\Lambda)\nonumber\\
&-&f_{n}(k_{C})+f_{n}(k_{F})\theta(k_{C}-k_{F})
+\left(f_{n}(k_{C})+f_{q}(k_{F})-f_{q}(k_{C})\right)\theta(k_{F}-k_{C})=0,\
\end{eqnarray}
with 
\begin{equation}
f_{x}(k)=\frac{\gamma_{x}g^{2}_{s x}\bar{\sigma}}{4\pi^{2}}\left(k \sqrt{k^{2}+(g_{s x}\bar{\sigma})^{2}}-(g_{s x}\bar{\sigma})^{2}
\ln\frac{k+\sqrt{k^{2}+(g_{s x}\bar{\sigma})^{2}}}{(g_{s x}\bar{\sigma})^{2}}\right),
\end{equation}
where we used the notation that $x=q,n$ stands for the quark and the nucleon terms, respectively.

The quark interaction term Eq.~(\ref{l-q}) introduces a new extra
parameter $g_{s q}$ which is fixed by choosing the quark mass at
vacuum $m_{q}(\sigma_{0})=300$ MeV, therefore we have $g_{s
q}=3.23$. The other parameters are determined for a given value of
deconfinement density $\rho_{C}$ (or associated Fermi momentum
$k_{C}$) in the same procedure described in section II, so as to
reproduce the empirical nuclear matter saturation point
$(E_{B}/A=-15.75~\text{MeV},\rho_{0}=0.148~ \text{fm}^{-3})$. Therefore, we
calibrate all parameters to nuclear matter properties. We choose again,
$M_{N}(\sigma_{0})=939$ MeV, $m_{v}=783$ MeV and $f_{\pi}=93$ MeV in
vacuum. Notice that, here the cutoff $\Lambda$ depends on the
deconfinement density or equivalently Fermi momentum $k_{C}$, see
Eq.~(\ref{eqr}). This is in fact a cutoff available for a quark in the
phase space measured relative to the critical deconfinement Fermi
momentum $k_{C}$. A density dependent cutoff was also invoked by Harada,
Kim and Rho \cite{h22} to study the chiral restoration in the context
of vector manifestation scenario of Harada and Yamawaki \cite{h6}.

In contrast to the case that transition to quark degrees of freedom is
not incorporated, here the cutoff can be taken larger and
cutoff-dependence is less pronounced.  In order to compare our result
with the previous section we take the cutoff given in table I
for set $B3$, $\Lambda=483.2$ MeV.  In table II, we show two sets of parameters determined
for the fixed cutoff $\Lambda=483.2$ MeV assuming that deconfinement take place at
density $\rho_{C}=3\rho_{0}$ and $1.5 \rho_{0}$ for parameter sets $C1$ and
$C2$, respectively. As we already pointed out the cutoff depends on
the deconfinement density, this means for a fixed cutoff one needs to adjust
other parameters in order to reproduce again given phenomenological
input, see table II.  
\begin{table}
\caption{The parameters $\Lambda$, $m_{s}^{2}$, $\lambda$ and $g_{vn}$ are shown for sets $C1$ and $C2$ for two values of the critical deconfinement density $\rho_{C}$. We assume a cutoff $\Lambda=483.2$ MeV and the coupling $g_{s q}=3.226$ (which yields constituent quark mass $300$ MeV) for all parameter sets. All parameter sets reproduce the empirical saturation point  $(E_{B}/A=-15.75~\text{MeV},
\rho_{0}=0.148~ \text{fm}^{-3})$.   
 The corresponding sigma mass $m_{\sigma}$ in the vacuum and the
 resulting in-medium nucleon mass $M_{N}^{\star}$ at saturation
 density are also given.}
\begin{ruledtabular}
\begin{tabular}{lll}
Parameter &set $C1$ & set $C2$ \\
\hline
$k_{C}$ (MeV) & 368.8 $(\rho_{C}=3\rho_{0})$ &292.7 ($\rho_{C}=1.5\rho_{0}$) \\
$m_{s}^{2} (\text{GeV}^{2})$ &0.48 &0.376  \\
$\lambda$ & 15 &20 \\
$g_{vn}$  & 6.35 & 6.54  \\
\hline
\hline
$m_{\sigma}(\bar{\sigma}_{0})$ (MeV) &815&807\\
$M_{N}^{\star}$(MeV)&772.2 &762.3\\
\end{tabular}
\end{ruledtabular}
\end{table}

In Fig.~3 left panel, we show the mean-value of scalar field as a
function of density for various parameter sets $B3$ and $C1,C2$ in the
presence of the vector-scalar interaction terms $g_{v}\neq 0$ given in
table II. For comparison, we also plotted again the corresponding
result for parameter set $B3$ given in table I (without the
deconfinement effect). It is seen that exactly at the critical
deconfinement density $\rho_{C}$, the slope of in-medium scalar-mean
field changes toward chiral restoration (for set $C2$ this is more
obvious). This is in contrast to pure nuclear matter case set $B3$
(see also Fig.~1 left panel) where chiral restoration seems to be in
conflict with nuclear matter properties.  On the right panel of
Fig.~3, we show the binding energy per baryon for various sets. It is
seen that the equation of state at very high density $\rho\approx
9\rho_{0}$ become softer than pure nuclear matter. At the
deconfinement density there is a pronounced kink which is more obvious
if the deconfinement occurs at lower density (see Fig.~3, set
$C2$). This is due to the fact that a sharp boundary is assumed
between the baryonic and the quark phases. In principle, instead of
the theta function in Eq.~(\ref{eqr}) one may employ a smooth
function. Note also that here the formation of Cooper pairs and
color-superconductivity is not taken into account \cite{h9}. It has
been shown that such effects might have important consequences for the
transition to quark matter within compact stars \cite{compact}. In
particular, diquark degrees of freedom seem to be essential for the
emergence of coexistence region between the hadronic and deconfined
phases \cite{di-rp}.

\begin{figure}[!tp]
       \centerline{\includegraphics[width=16 cm] {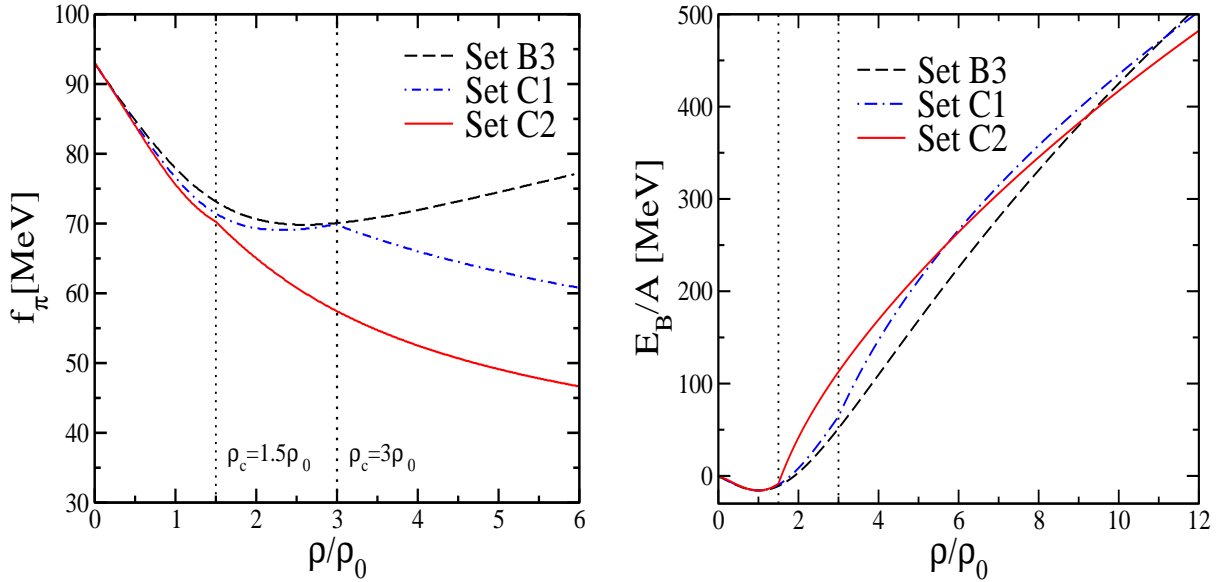}}
       \caption{On the left panel: the scalar mean-field values with
       respect to density $\rho/\rho_{0}$ (with nuclear matter density
       $\rho_{0}=0.15 fm^{-3}$) is shown when $g_{v}\neq 0$ for various parameter
       sets $B3$ and $C1,C2$ given in table I and II, respectively. On
       the right panel: we show the binding energy per baryon $E_{B}/A$ as a
       function of density $\rho/\rho_{0}$ for
       the same sets of parameters. The vertical dotted lines
       show the positions of the critical deconfinement density $\rho_{C}$.}
\end{figure}

Next, we switch off completely the vector-scalar interaction term
$g_{v}=0, $ (i.e. $\mathcal{R}=0$ in Eq.~(\ref{eqr})) above the
deconfinement density $\rho_{C}$. In Fig.~5, we show the mean-value of
scalar field as a function of density for parameter sets $C1$ and $C2$
when $g_{v}=0$ for $\rho>\rho_{C}$. If the deconfinement take place at higher density the
chiral phase transition will be stronger and coincide with the 
deconfinement critical density. In this case the chiral phase
transition is first order. While for a soon deconfinement (for
example, parameter set $C2$) the chiral symmetry is partially restored
at the deconfinement density and full chiral restoration postpones at
relatively higher density than the deconfinement density
$\rho_{C}$. It is interesting to note that the soon deconfinement
(parameter set $C1$) is not compatible with Brown-Rho scaling
\cite{br},
\begin{equation}
\frac{f_{\pi}(\sigma)}{f_{\pi}(\sigma_{0})}\approx \frac{M_{N}(\sigma)}{M_{N}(\sigma_{0})}\approx \frac{m_{v}(\sigma)}{m_{v}(\sigma_{0})}, 
\end{equation}
since we have $m_{v}(\sigma)=0$ above the deconfinement when
$g_{v}=0$, but as it can be seen from Fig.~5 for parameter set $C1$,
the mean-value of scalar field $f_{\pi}$ above the deconfinement is
partially restored and is not zero. If we assume that Brown-Rho
scaling is valid at various density, we can put a constraint from
below on the onset of critical deconfinement density $\rho_{C}$ in our
minimal Lagrangian model. We found that for the deconfinement density
$\rho_{C}\geq 2\rho_{0}$, we have $f_{\pi}(\rho_{C})=0$ and
$m_{v}(\rho_{C})=0$ at the same time consistent with Brown-Rho scaling and vector
manifestation scenario which is considered to be a generic feature of
effective field theory matched to QCD \cite{h6}. Therefore, for
$\rho_{C}\geq 2\rho_{0}$, the chiral restoration is first order and
coincide with the critical deconfinement density $\rho_{C}$.

\begin{figure}[!tp]
       \centerline{\includegraphics[width=8 cm] {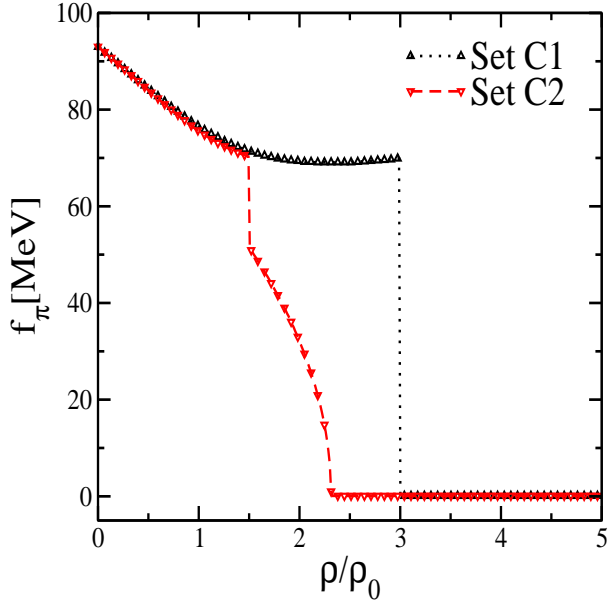}}
       \caption{ We show the scalar mean-field values $f_{\pi}$ as a
       function of density $\rho/\rho_{0}$ for parameter sets $C1$ and
       $C2$ given in table II when the vector-coupling $g_{v}$ is zero
       above the critical deconfinement density $\rho_{C}$.}
\end{figure}

\section{conclusion}
 We investigated the implication of the baryonic Dirac-sea in the presence
 of an explicit ultraviolet cutoff in the nuclear matter equation of
 state within the L$\sigma$N model.  In this way, the model becomes
 non-renormalizable and one can avoid the unnaturalness problem
 \cite{h11}.  We investigated the parameter space of the model as a
 function of cutoff which reproduces the empirical nuclear matter
 properties. Note that the empirical compression modulus was not
 used in the fitting procedure.  We showed that a filled Dirac sea of
 baryons can produce a strong attraction in nuclear matter. It softens
 the equation of state and it generates a vacuum with dynamically broken symmetry even
 without a presence of a negative mass term in the linear sigma
 model. This also clarified a possible connection between the chiral
 L$\sigma$N model and bosonized version of the NJL model in a nuclear
 matter medium.

Although the L$\sigma$N model in the presence of the vector-scalar
interaction can describe the nuclear matter saturation, it is unable
to accommodate the chiral restoration phase transition with
$f_{\pi}=0$. The role of vector-meson is very subtle in this model. On
one hand it is needed in order to reproduce the empirical saturation
properties, but it works against the chiral restoration at high density.
In order to study the implication of the first-order deconfinement
effect in the equation of state and chiral phase transition, we simulated
the deconfinement effect by changing the active degrees of freedom from
nucleons to quarks above the critical deconfinement density. At the
moment, we do not know yet the order of a possible deconfinement
transition and its critical density. However, there are some
indications that it might be first order and takes place at about
several times the nuclear matter density \cite{h9}. Here, we
considered the implication of different choices of the critical
deconfinement density. We investigated the role of the vector-meson
and the deconfinement effect in describing the chiral phase transition
when our model parameters are calibrated to the nuclear matter
properties. The nature of the chiral phase transition is sensitively
dependent on the presence of the vector-scalar interaction.  The
vector-meson has been reported by many authors to be important for
realization of the chiral phase transition \cite{bub}.  We showed that
if the vector-meson is completely decoupled from the quark matter
$g_{v}=0$ above the critical deconfinement density, chiral phase transition can take
place and its position depends on the given critical deconfinement
density. We showed in order to be consistent with Brown-Rho scaling
\cite{br} and vector-realization \cite{h6} based on hidden-local
symmetry, the critical deconfinement density must be bigger than
$2\rho_{0}$ where $\rho_{0}$ is the nuclear matter density. Then the
chiral restoration is first order and is coincident with the deconfinement
density\footnote{Notice that in contrast with the finite density where
Lattice simulation is plagued with sign problem, lattice calculation
has been very fruitful for finite-temperature (at zero baryon density)
studies \cite{la1}. An analysis of the lattice simulation shows that chiral symmetry
restoration and deconfinement phase transition occurs at the same
temperature \cite{la1,la2}. It has been also shown that the vector coupling is small
compared to the scalar coupling at high temperature \cite{la3}.}. 

It is of interest to extend scheme presented in this paper to finite
temperature and also study the implication of this nuclear-quark equation of state on neutron star.

\section*{Acknowledgements}
The author is very grateful to Hans J. Pirner for very fruitful discussions and to J\"org Raufeisen for careful reading of
the manuscript and comments. This work was supported by the Alexander
von Humboldt foundation.

\end{document}